\documentclass[prd,nopacs,nokeys,nofootinbib]{revtex4}
\usepackage{amsfonts}
\usepackage{amsmath}
\usepackage{amssymb}
\usepackage{amsmath}
\usepackage{bbm}
\usepackage[utf8]{inputenc}
\usepackage[caption=false]{subfig}
\usepackage{tensor}
\usepackage{slashed}
\usepackage[centertableaux ]{ytableau}
\usepackage{hyperref}
\usepackage{natbib}
\usepackage{braket,mleftright}
\usepackage{graphicx}
\usepackage{cleveref}
\usepackage{tikz}
\usepackage{tikz,ifthen}

\usetikzlibrary{trees}
\usetikzlibrary{decorations.pathmorphing}
\usetikzlibrary{decorations.markings}
\tikzset{beamerprimary/.style={structure.fg, thick}}
\tikzset{beamersecondary/.style={structure.bg, thick}}
\tikzset{boson/.style={draw=structure.fg,decorate, decoration={snake}},
    gauge/.style={decorate, decoration={snake} },
    fermion/.style={postaction={decorate},
         decoration={markings,mark=at position .55 with {\arrow{>}}}},
    fermionloop/.style={postaction={decorate},
        },
    gluon/.style={decorate,
        decoration={coil,amplitude=4pt, segment length=5pt}},
    scalar/.style={dashed},
    graviton/.style={double}
}

\begin{document}

\title{Morphology and numerical characteristics of epidemic curves for SARS-Cov-II using Moyal distribution}
\author{Jos\'e de Jes\'us Bernal-Alvarado}
\email{bernal@ugto.mx}
\affiliation{Physics Engineering  Department,  Universidad de Guanajuato\\
Lomas del Bosque 103, Fraccionamiento Lomas del Campestre, 37150, Le\'on,
Guanajuato, M\'{e}xico}

\author { David Delepine}
\email{delepine@ugto.mx}
\affiliation{Physics Department,  Universidad de Guanajuato\\
Lomas del Bosque 103, Fraccionamiento Lomas del Campestre, 37150, Le\'on,
Guanajuato, M\'{e}xico}


\begin{center}

\today

\begin{abstract}
 In this paper, it is shown that the Moyal distribution is an excelent tool to study the SARS-Cov-II (Covid-19) epidemiological associated curves and its propagation. The Moyal parameters give all the information to describe the form and the impact of the illness outbreak in the different affected countries and its global impact.  We checked that the Moyal distribution can accurately fit the daily report of {\it{new confirmed cases of infected people}} (NCC) per country, in that places where the contagion is reaching their final phase, describing the beginning, the most intense phase and the descend of the contagion, simultaneously . In order to achieve the purpose of this work, it is important to work with a complete and  well compilated set of the data to be used to fit the curves. Data from European countries like France, Spain, Italy Belgium, Sweden, United Kingdom, Denmark and others like USA and China, have been used. Also, the correlation between the parameters of the Moyal distribution fitting and the general public health measures imposed in each country, have been discussed. A relation between those policies and  the features of the Moyal distribution, in terms of their parameters and critical points, is shown; from that, it can be seen that the knowledge of the time evolution of the epidemiological curve, their critical points, superposition properties and rates of the rising and the ending, could help to find a way  to estimate the efficiency of social distancing measures, imposed in each country, and anticipate the different phases of the pandemic.
\end{abstract}

\end{center}
\maketitle

\section{Introduction}

Given the development of the pandemic, caused by the SARS-CoV-2 (Covid-19) outbreak, it is urgent and a priority to take decisions on public health policies and applicate them immediately.
Data analysis applied to epidemiological records can provide immediate information on the trend and progress of the contagion. The statistical distribution of the large amount of affected people, NCC , as well as the hospitalized and deceased individuals, shows a trend and a regular pattern common to most of the cases, especially when cases by country are studied. Particular variations  in the statistics of the NCC, along the time, can be clasificated in a finite number of categories, all of them, based in the same mathematical model. These data are available today, thanks to the large number of websites of government institutions that make such information available to the public. In this paper the data from \cite{cdceurope} are used. Shortly after the outbreak appeared in the Chinese city of Wuhan, specific theoretical works for this problem started to be published \cite{DellaMorte:2020wlc,Anacleto:2020ull,Cacciapaglia:2020mjf}, trying to adapt and generalize the SIR\cite{sir1,sir2,sir3,sir4,sir5,sir6,sir7,sir8,sir9,sir10} or SEIR model\cite{seir1,seir2,seir3,seir4,seir5} using the information that was available (or inferred) at the time, related to the parameters of the system of simultaneous differential equations \cite{brauer}. On the other hand, the limited knowlegde on this suddenly and complex phenomenon, in the geometric and numerical details of the epidemiological curve, has been a  barrier for anticipating the degrees of freedom in the analisys of the statistical distributions of affected people, as a function of time, as well as a dare to know their symmetries, critical points, growth constants, extinction times, etc. So, alternative methods based on empirical methods\cite{e1,e2,e3,e4,e5,e6,e7,e8,e9,e10,e11} are needed to predict and to measure the effects of  pulblic health policy to contain Covid-19. With the presented methodology, it is not only possible to characterize the evolution of the phenomenon, but also to identify anomalies in its behavior and sanitary risks that could menace the population (for instance, new outbreaks), offering a systematic way for planning and  implementing  of political, sanitary and economic strategies that will serve as a guide for the removal of the quarantine, the return to confinement or to anticipate a new outbreaks. Also, this is an effective methodology to measure the start and the ending times of the epidemy  phases.

\section{Metodology}

Lets consider, as a case study, the number of NCC per day. The outbreak in Italy was one of the first to develop and strongly affect a significant number of its inhabitants. The progress per day is very well documented and, at this time, it is in the extinction stage. Given the characteristics of the distribution of data over time, the best way to mathematically describe the number of NCC per day is using the Moyal distribution $\Lambda (t,A,\mu,\beta) $, \cite{Moyal:2009xna} which is also known to be a very accurate approximation of the Landau distribution \cite{Landau:1941vsj}

\begin{equation}
\Lambda(t)=  \frac{A}{\beta \sqrt{2\pi}} e^{-\frac{1}{2}\left[e^{-\left( \frac{t-\mu}{\beta} \right)}+\left( \frac{t-\mu}{\beta}\right)\right]}
\end{equation}
whose derivative can be expresed in the form of a generalized exponential growth of the form:
\begin{equation}
\frac{d\Lambda(t)}{dt}=r(t)\Lambda(t)
\end{equation}
where $\it{r(t)}$ is the ratio of growing with units of $time^{-1}$(in this work the time unit is $day$):

\begin{equation}
r(t)=\frac{1}{2\beta} \left[e^{-\frac{t-\mu}{\beta}}-1\right]
\end{equation}
The theoretical interpretation of the Moyal distribution parameters is straighforward: i)  The global factor and area under the curve, $A$, correspond to the total of the affected individuals by Covid-19, since the beggining of the outbreak up to their ending. ii) from $r(t)\Lambda(t)=0$, $\Lambda(\mu)$ is the maximum number of NCC at the day $t=\mu$; iii) $\beta$ is a parameter, with units of $day$, that uniquely characterizes the growth and extinction times of the epidemic outbreak. In a general sense, it is the time constant of the epidemic. It is remarkable that the affectation, in the different  categories of the victims: Infected, hospitalized and deceased, have the same behavior and similar time constants for a particular outbreak. Also, it is possible to identify the temporal delay between the maximum values of the each distribution, from which the average recovery or death times can be quantified.

\section{Mathematical model}

The applied mathematical model to the data analysis is based on the hypothesis of an exponencial behaviour of the COVID-19, where the time variation of the function is proportional to the number of cases at time t, modulated by a time function. The cumulative distribution associated to Moyal distribution is given by:

\begin{equation}
\int_0^t \Lambda(p)dp=Erfc\left(\frac{ e^{\frac{t-\mu}{\beta}}}{\sqrt{2}}\right) 
\end{equation}
where $Erfc(y)$ is the well-known complementary gauss error function:

\begin{equation}
Erfc(y)=\frac{2}{\sqrt{\pi}}\int_y^{+\infty} e^{-u^2}du
\end{equation}

The error function is very close to the logistic function which is usually used to describe pandemic propagation\cite{e2,e3}. But in our case the derivative of the logistic function is too symmetric to describe well the tails of the new dialy infected cases for Covid-19 as observed in countries where the Covid-19 pandemy is slowly down and passed its highest point. Also the error function appears in solution of lineal difusion equation with constant coefficient of diffusion. 
It is interesting to get the critical points of the Moyal distribution. As said before, the maximum value of the Moyal distribution is obtained at $t=\mu$. But it is also possible to get the inflexion point studying the second derivative:
So, one immediately gets:
\begin{equation}
\frac{d^2\Lambda(t)}{dt^2}=\Lambda(t)\left[\frac{dr(t)}{dt}+\left[r(t)\right]^2\right]
\end{equation}

 the inflection points are the solution of the equation:

\begin{equation}
\left[\frac{dr(t)}{dt}+\left[r(t)\right]^2\right]=0
\end{equation}

substituting $r(t)$,

\begin{equation}
\left(e^\frac{t-\mu}{\beta}\right)^2+4\left(e^\frac{t-\mu}{\beta}\right)+1=0
\end{equation}

that is $t=-\beta Ln(2\pm\sqrt{3})+\mu$. So, the inflection points $B_1$ and $B_2$ are given by:
\begin{equation}
B_1=-1.317\beta+\mu, \ \ \ B_2=+1.317\beta +\mu.
\end{equation}

\begin{figure}
	\centering
	\includegraphics[width=0.7\linewidth]{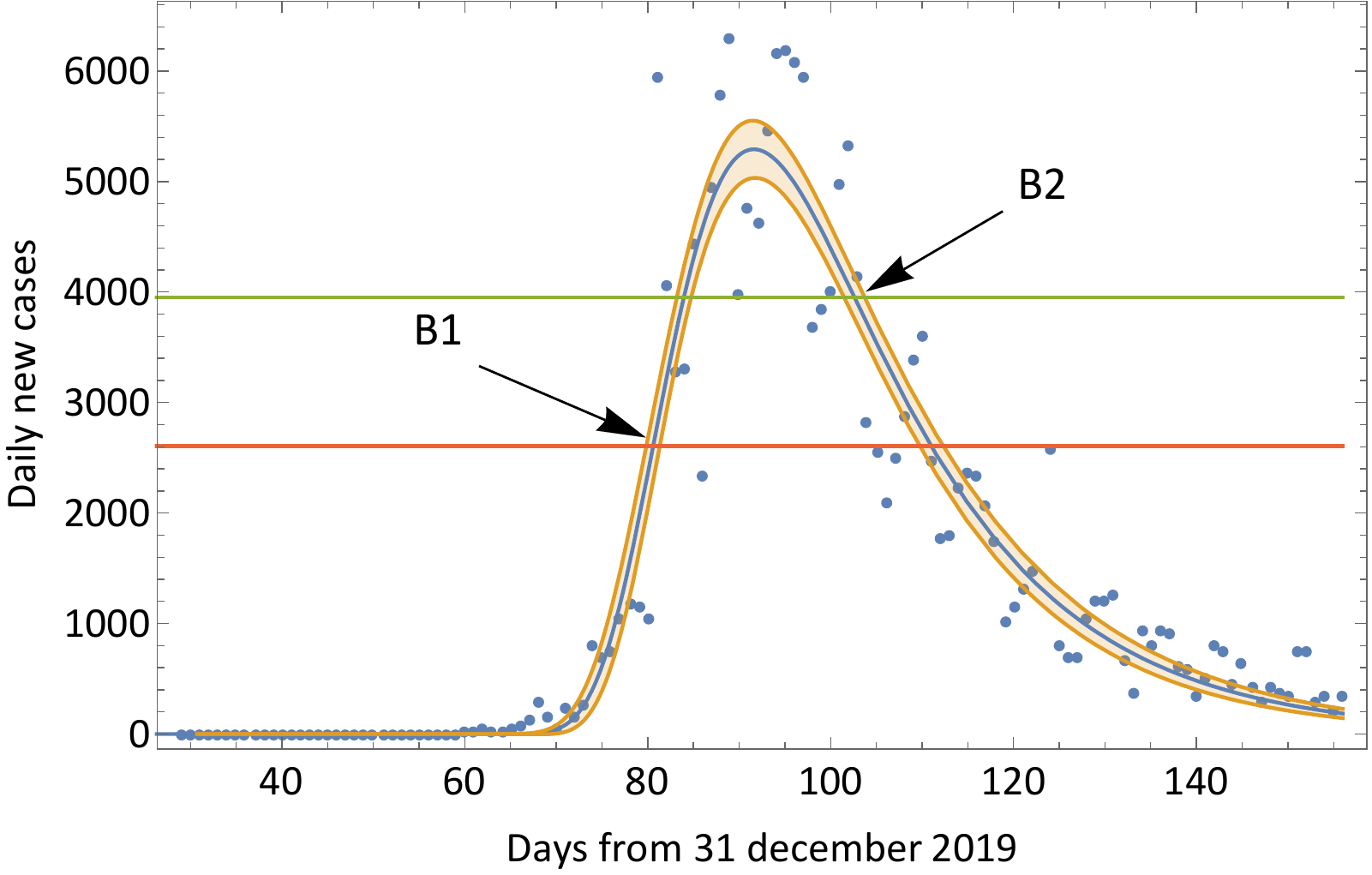}
	\caption{Mathematical model applied to Germany. The inflection points mark one half of the maximun number of daily NCC: $\Lambda(B_1)=0.493\Lambda(\mu)$, while  $\Lambda(B_2)=0.746\Lambda(\mu)$, pointing out a descence of one quarter from the top of the curve.}
	\label{Alemania}
\end{figure}

Obviously, the four parameters ($A,\beta,\mu, B_1$ and $B_2$) obtained through the fit of $\Lambda(t)$  to experimental data are scale invariant, mark the escential points of the epidemiological curve and define the evolution of the epidemy for all time; for instance $\Lambda(B_1)=0.493\Lambda(\mu)$. In figure \ref{Alemania}, the model is applied to data from Germany.

\begin{figure}
	\centering
	\includegraphics[width=0.7\linewidth]{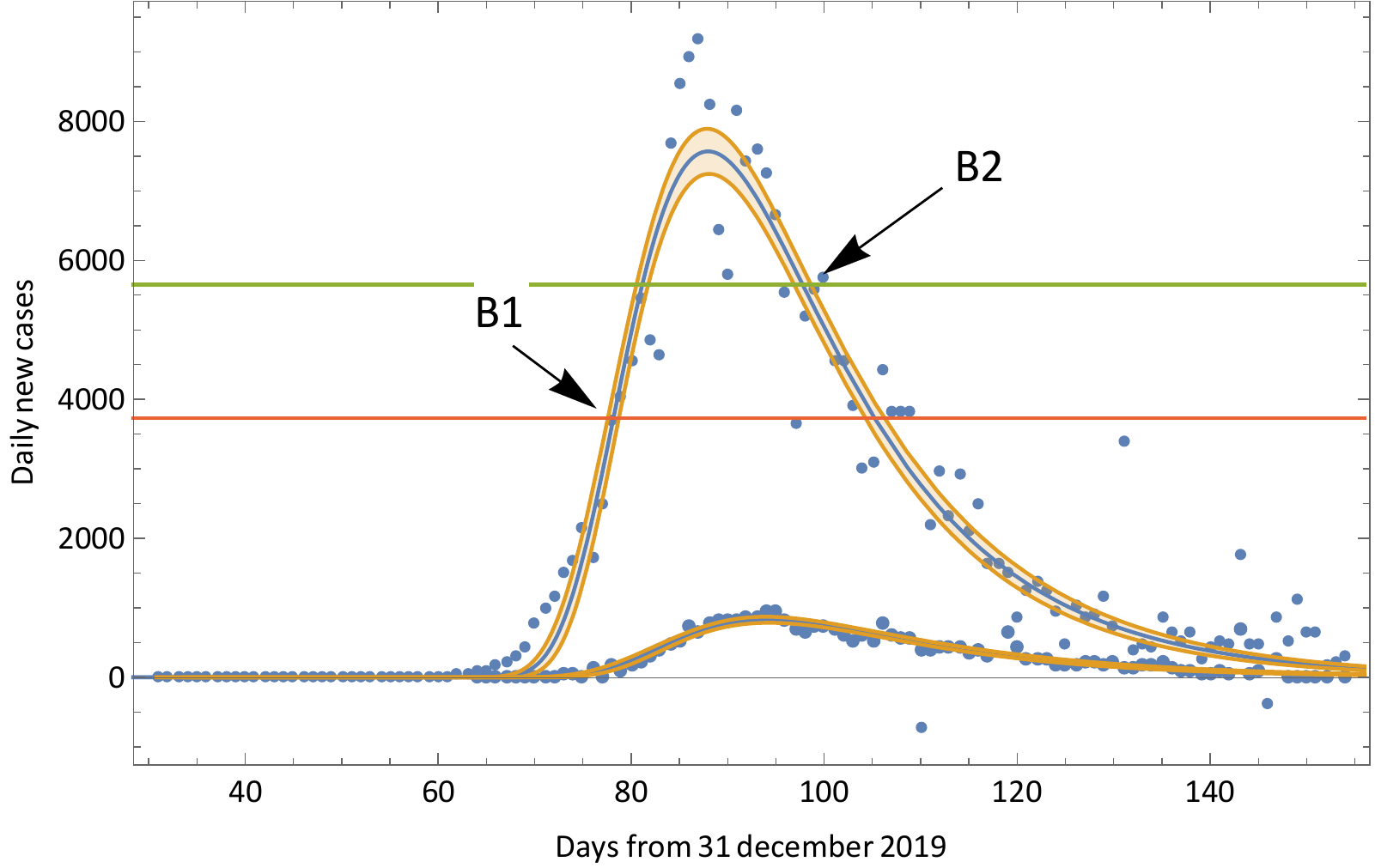}
	\caption{Distribution of the daily NCC (higher curve with $B_1$ and $B_2$ pointing out the inflection points) and deceased persons in Spain.}
	\label{Spain}
\end{figure}


\section{Results}

\begin{table}
\caption{ Fitting parameters of the Moyal function on the daily NCC in different countries, assuming day 1 is the 31st. december 2019.\label{fit}}
{%
\vspace{10mm}
\begin{tabular}{|c|c|c|c|c|}
\hline
countries & $\mu$ &  $\beta$ & A& total of reported cases \\
& & & &  (date:  3 june 2020) \\
\hline
\hline

Italia & 88.22 (0.32) & 10.83 (0.24)& 240320 (4295)& 234013 \\
Spain& 87.97 (0.32) & 7.45 (0.24) & 232961 (6033) & 241000 \\
Luxemburg& 79.21 (0.47) & 5.49 (0.35) & 3881 (202)& 4020 \\
Danemark & 95.58 (0.99)& 12.62 (0.78) & 12442 (609)& 11811 \\
Belgium & 98.68 (0.66) & 9.99 (0.50) & 61048 (2450)& 58767 \\
Germany & 91.61 (0.41)& 8.36 (0.31) &182839 (5451) & 184597 \\
Sweden & 123.23 (2.10) & 26.29 (1.72) & 65132 (3824)& 41833 \\
France & 94.47 (0.62)& 8.55 (0.46)&148229 (6525) & 152444\\
United Kingdom & 110.07 (0.77)& 14.81 (0.65) & 333511 (11345) & 282000\\
China & 39.12 (0.73) & 5.09 (0.54) & 85521 (7395)& 83022 \\
USA & 112.48 (0.77) & 18.64 (0.66) & 2430429 (68704) & 1910735 \\
\hline
\end{tabular}}
\end{table}

The growth pattern is not only qualitatively similar in the different epidemiological outbreaks, following an asymmetric bell shape, furthermore, the numerical parameters that characterize the duration of the epidemic and the relative location of its maximum point, are determined by the value of $\beta$: $\Delta t=2(1.317)\beta$ is the lapse to reach the maximum of daily new cases (for any category) starting from $0.006\Lambda(\mu)$. On the other slope of the curve,  $\Delta t=5(1.317)\beta$ is the lapse to diminish the daily NCC, from the maximum value, to $0.06\Lambda(\mu)$. Figure \ref{Spain} shows the main qualitative and quantitative characteristics of the epidemiological curve: the asymmetric bell shape according to the Moyal distribuion, in which the position of the maxima clearly shows the average time between the maximum of confirmed cases and the maximum number of deaths. Likewise, a fundamental property of the asymmetry of the curve is shown, which is inherent to the mathematical function that describes it: the beginning of the growth of the cases at $t= -2(1.317)\beta+\mu$ , prior to the maximum, as well as the extinction of the curve around $t= 5(1.317)\beta+\mu$.

From table \ref{fit}, it is interesting that the countries can be classified, depending of the value of $\beta$,  as shown in table \ref{fitbeta}. Contrasting this table with the general public health policy applied in each country to fight against the Covid-19. European countries like Spain, Italy, Belgium, Germany and France have globally applied the same kinds of measures to restraint the propagation of the illness. Sweden is the main European country who decided not to apply quarentine. United Kingdom started with a kind of policy similar to Sweden, nevertheless, they eventually changed to quarentine measures, as continental european countries like France and Spain did. In USA, the decision to implement a quarentine or not was given to the states government, but under he pressure from federal governement  to limit the measures as much as was possible. China imposed the most strict confinement measures  and its very low value of $\beta$ is an expression of the result of this policy.

In the next subsection, we shall discuss with more details the situation in countries like Sweden, USA and United Kingdom to illustrate the effects of general public health policy in the epidemiological curve describing Covid-19 propagation. 

\begin{table}
\caption{ Classificacion of countries according to  their $\beta$ value.\label{fitbeta}}
{%
\begin{tabular}{|c|c|c|c|}
\hline
$\beta <6$ & $ 6<\beta<12$  & $12<\beta<15 $  & $\beta >15$ \\
\hline
China& Spain & Danemark & USA \\
Luxemburg & Italia&United kingdom& Sweden\\
 & Germany& &  \\
 & Belgium& & \\
 & France& & \\
\hline
\end{tabular}}
\end{table}

\subsection{Sweden case}

In the case of Sweden, the daily NCC are very dispersed after reaching  its maximun around $t=100$. Nevertheless, despite the fluctuations, the average trend is not going down; instead of that, the number of cases agrees with a new modelization strategy: a suitable fitting function $\Lambda_T (t)$, must describe the appearence of new outbreaks through the sum of differen Moyal distribution, one for  each independent outbreaks. Of course, once a new contribution is added, in the sum of Moyal functions, new degrees of freedom  are added to the fit and this could increase the fitting instability and difficult to do the algorithm converging. For simplicity, the sum of two independent Moyal distribution is done, this can be written as follow:

\begin{equation}
\Lambda_{T}(t)=\Lambda_1 (t,A_1,\beta_1,\mu_1)+\Lambda_2 (t,A_2,\beta_2,\mu_2) \label{X}
\end{equation}

\begin{figure}
	\centering
	\includegraphics[width=0.7\linewidth]{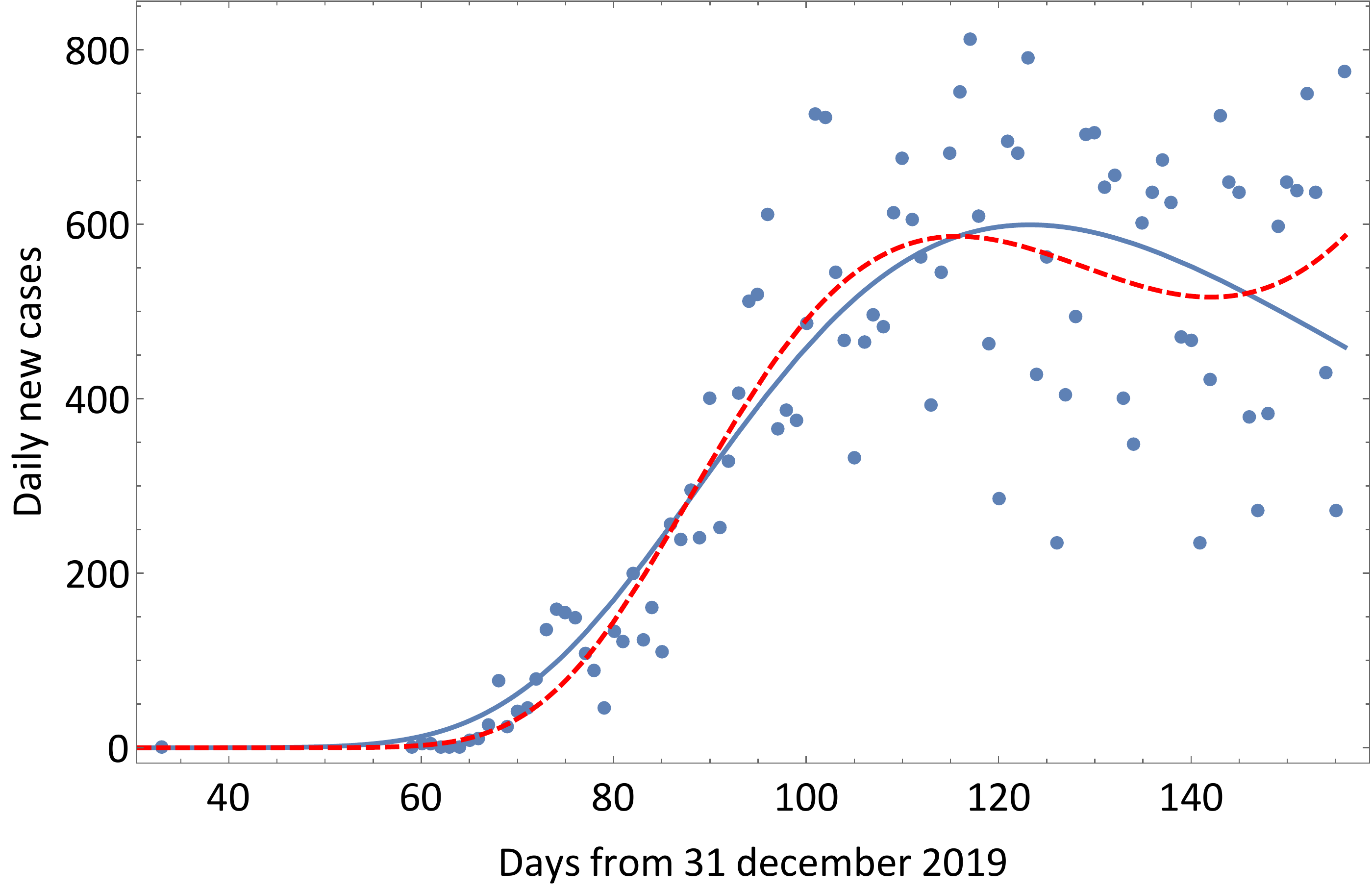}
	\caption{Distribution of the new infected cases in Sweden. The dashed line is a linear combination of two Moyal distribution and the solid line corresponds to only one Moyal distribution.}
	\label{sweden}
\end{figure}

 In the figure \ref{sweden}, the situation in Sweden is shown first using a Moyal distribution (the solid line) and, as expected, the tail go down slowly. The dashed line represents the sum of two Moyal distributions and at present time and in that case,  the best fit  prefer a trend to increase the  average values, in the number of daily NCC, which could mean that a new outbreak is taking place at this time. Meanwhile new avalaible data are published, both options must be followed.

\subsection{ United Kindgom case}

\begin{figure}
	\centering
	\includegraphics[width=0.7\linewidth]{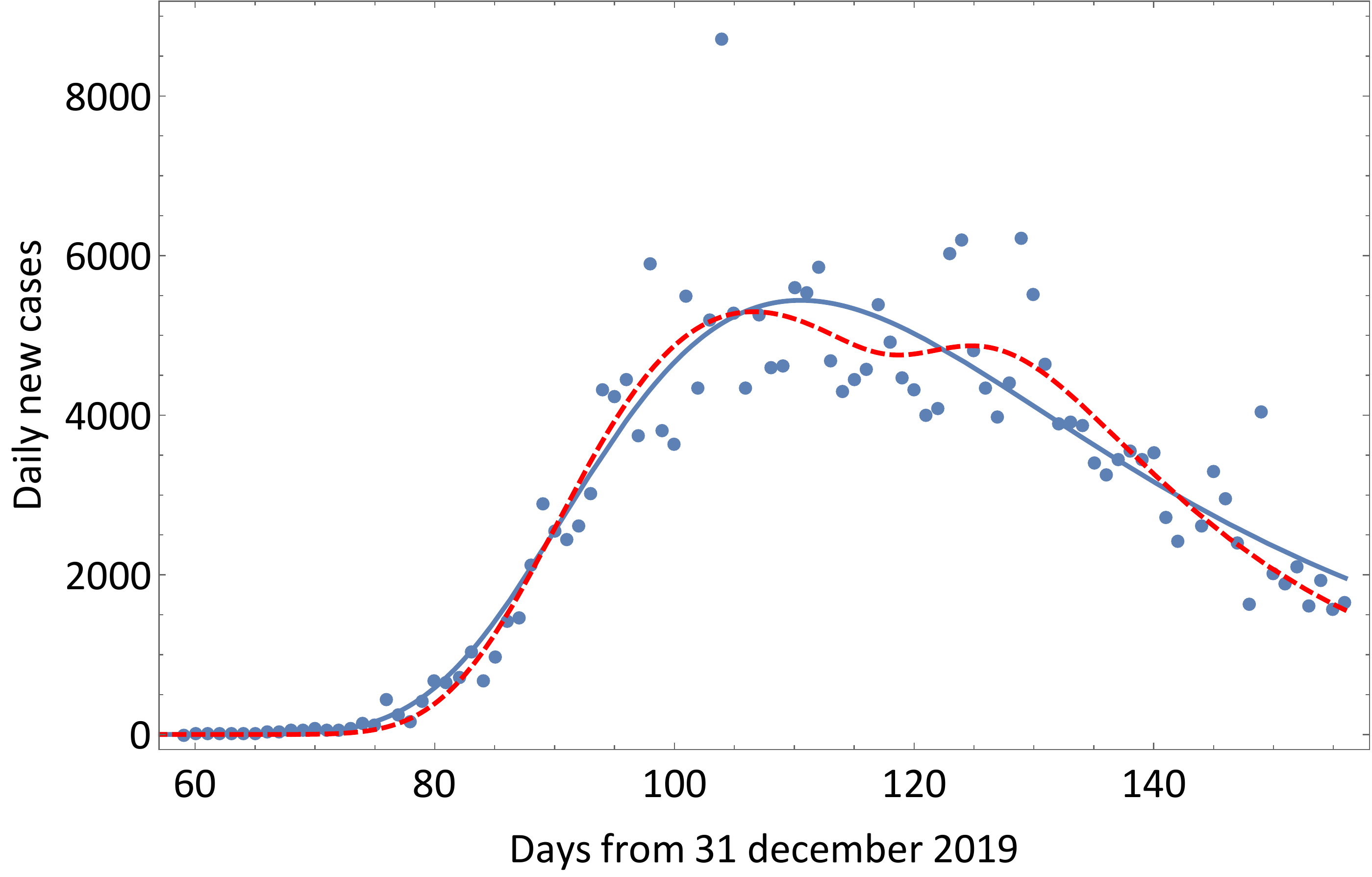}
	\caption{Distribution of daily NCC in United Kingdom. The dashed line is the sum of two Moyal distribution and the solid line corresponds to  one Moyal distribution.}
	\label{uk}
\end{figure}

An interesting case, to prove the efficacy to fit a function $\Lambda_T (t)$ , as described in equation \ref{X}, is the anslysis of the United Kingdom data. This case have a small but clear bump just after reach the maximum of the daily NCC.  The combination of two Moyal functions, is consistent with the assumption of two independent outbreaks, that contribute to the total amount of cases. In figure \ref{uk}, the solid line corresponds to the Moyal distribution with the parameters given in table \ref{fit}. The dashed line correspond to the sum of two Moyal distribution.

\begin{table}
\caption{ Fitting parameters, using two Moyal distributions to fit the daily NCC data for United Kingdom.}
\label{fitusa}
\begin{tabular}{|c|c|c|c|c|c|}
\hline
$ \mu_{1} $ &  $\beta_{1} $  &  $A_1 $  &  $\mu_{2} $ &  $\beta_{2}$ & $A_2$  \\
\hline
\hline
106.6 (1.66) & 12.54 (1.21)& 274458 (27790)& 130.61 (1.74)& 6.69 (2.03) & 39714 (21139) \\
\hline
\end{tabular}
\end{table}

\subsection{ USA case}

\begin{figure}
	\centering
	\includegraphics[width=0.7\linewidth]{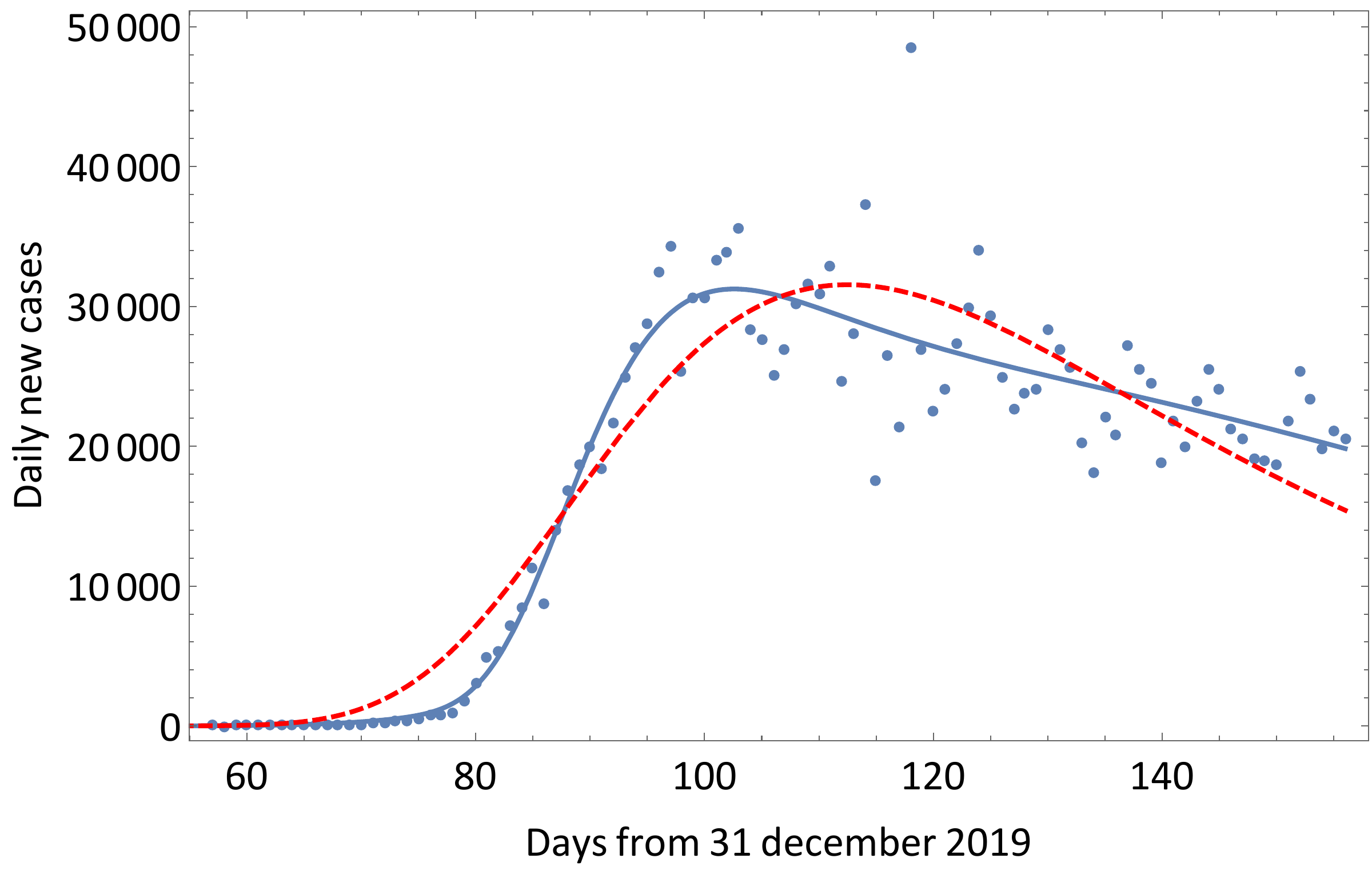}
	\caption{Distribution of the new infected cases in USA. The think line is the sum of two Moyal distribution and the dashed line corresponds to onyl one Moyal distribution.}
	\label{usa}
\end{figure}

In the USA case, the  outgoing tail takes  an unusual longer time to decrease, nevertheless, it could be considered according to one Moyal Distribution, nevertheless, this occurr in first approximation; a better data description is achieved with the sum of two Moyal distribution. In figure \ref{usa}, it can be seen the results when two Moyal distribution are used. the dashed line corresponds to the fitting of a single Moyal distribution.  The parameters values are given in table \ref{fit}. The solid line corresponds to the sum of the two Moyal distribution. The parameters values of the two Moyal distribution are given in table \ref{fitusa}. This could be interpreted from the epidemilogical point of view because the fact that USA has several different outbreak of the Covid-19, each of them with different time of starting and  parameters. It is interesting that the first Moyal distribution parameters are very closed to the New York Covid outbreak  where the maximum of NCC took place around the 7th April and the number of infected persons are around 380,000, within one error sigma of our first Moyal distribution parameters.

\begin{table}
\caption{ Fitting parameters of the Moyal function using two distributions to fit  the USA data.}
\label{fitusa}
\begin{tabular}{|c|c|c|c|c|c|}
\hline
$ \mu_{1} $ &  $\beta_{1} $  &  $A_1 $  &  $\mu_{2} $ &  $\beta_{2}$ & $A_2$  \\
\hline
\hline
98.54 (1.99) & 8.55 (1.79)& 797664 (355417) & 139.80 (12.90) & 28.18 (4.84) & 2313510 (411043) \\
\hline
\end{tabular}
\end{table}

\section{Conclusion}
 In this paper, we have shown that the Moyal distribution is an excelent tool to study the Covid-19 epidemiological curves and its propagation. The Moyal parameters give all the information to describe the form and the impact of the propagation of the illness and its global impact on the country. Also it has been shown that the resutls of public health policy measures can be evaluated through the $\beta$ parameter. Clearly, this analisis could be extended to more countries with good statistics on daily new cases, in order to applicate, validade or generalize our hipothesis. A very interesting point is that it is possible to use this tool even before to reach the maximum point of new cases, in order to predict the effects of social measures imposed to restrain the propagation of the Covid-19. 

\begin{acknowledgments}
We acknowledge financial support from CONACYT and SNI (M\'exico).D. D. is  grateful to Conacyt (M\'exico) S.N.I. and Conacyt project (CB-286651), DAIP
project (Guanajuato University), PIFI (Secretaria de Educacion
Publica, M\'exico) for financial support.
\end{acknowledgments}


\end{document}